\documentclass[twocolumn,pra,aps,superscriptaddress, longbibliography]{revtex4-1}
\usepackage[T1]{fontenc}
\usepackage[latin9]{inputenc}

\usepackage{amsmath}
\usepackage{amssymb}
\usepackage{xcolor}
\usepackage{bbm}
\usepackage{mathrsfs}
\usepackage{physics}
\usepackage{float}
\allowdisplaybreaks
\usepackage{graphicx}

\makeatletter
\renewcommand*{\fnum@figure}{{\normalfont\bfseries \figurename~\thefigure}}
\makeatother

\usepackage[colorlinks=true]{hyperref}  
\hypersetup{
    bookmarks=true,         
    unicode=false,          
    pdftoolbar=true,        
    pdfmenubar=true,        
    pdffitwindow=false,     
    pdfstartview={FitH},    
    pdftitle={Euler Andreev Reflection},    
    pdfauthor={Morris, Bouhon, and Slager},     
    pdfsubject={},   
    pdfcreator={},   
    pdfproducer={}, 
    pdfkeywords={} {} {}, 
    pdfnewwindow=true,      
    colorlinks=true,       
    linkcolor=blue, 
    citecolor=blue,        
    filecolor=magenta,      
    urlcolor=blue,           
	breaklinks=true
}

\definecolor{orange}{rgb}{1,0.5,0}

\newcommand{\bea}{\begin{eqnarray}}
\newcommand{\eea}{\end{eqnarray}}

\renewcommand{\figurename}{Figure}

\newcommand{\ii}{\mathrm{i}}
\newcommand{\ee}{\mathrm{e}}
\newcommand{\EF}{E_{\text{F}}}
\DeclareMathOperator{\sgn}{sgn}

\begin{document}
\title{Andreev reflection in Euler materials}
%

\author{Arthur S. Morris}
\affiliation{TCM Group, Cavendish Laboratory, University of Cambridge, J. J. Thomson Avenue, Cambridge CB3 0HE, United Kingdom}
\author{Adrien Bouhon}
\affiliation{TCM Group, Cavendish Laboratory, University of Cambridge, J. J. Thomson Avenue, Cambridge CB3 0HE, United Kingdom}
\author{Robert-Jan Slager}
\affiliation{TCM Group, Cavendish Laboratory, University of Cambridge, J. J. Thomson Avenue, Cambridge CB3 0HE, United Kingdom}

\date{\today}
\begin{abstract}
Many previous studies of Andreev reflection have demonstrated that unusual effects can occur in media which have a nontrivial bulk topology. Following this line of investigation, we study Andreev reflection in topological Euler materials by analysing a simple model of a bulk node with a generic winding number $n\geq 0$. We find that the magnitudes of the resultant reflection coefficients depend strongly on whether the winding is even or odd. Moreover this parity dependence is reflected in the differential conductance curves, which are highly suppressed for $n$ even but not $n$ odd. This gives a possible route through which the recently discovered Euler topology could be probed experimentally.
\end{abstract}

\maketitle

\section{Introduction}

The study of topological materials is currently a very active field of research. Investigations in this area range from broad theoretical analyses to direct experimental investigation of the properties of particular materials. Recent advancements in the characterization of topological band structures using symmetry eigenvalue methods have been significant \cite{Kruthoff2019, Po_2017, Bradlyn_2017, Slager_2013,Scheurer2020, volovik2018investigation, Bouhon2018, Ft1, Slager2019, Fu2011}. However, there is growing interest in multi-gap topological phases~\cite{Bouhon2020} since they generically cannot be explained within this paradigm. 

A prominent example entails Euler topology~\cite{Bouhon2019, Ahn2019, Unal2020, Bouhon2018,ChiralHeirs}, which depends on the conditions between multiple gaps in the band structure of a $\mathcal{C}_2\mathcal{T}$ or $\mathcal{PT}$ symmetric material. Under these conditions, band nodes in two-dimensional materials carry non-Abelian `frame charges' in momentum space~\cite{wu2019non, Bouhon2019,Sjoqviist_2004}, akin to $\pi$ disclination defects in bi-axial nematics~\cite{Beekman20171,Liu2016,volovik2018investigation}; phases with a non-trivial Euler topology can be formed by braiding such degeneracies between successive bands \cite{wu2019non,Ahn2019,Bouhon2019,ChiralHeirs}. The ability of such nodes lying in a patch of the Brillouin zone, $\mathcal{D}\subseteq\text{BZ}$, to annihilate is encoded in the  $\mathbb{Z}$-valued Euler class invariant $\chi$, which is the real counterpart of the Chern number of a complex vector bundle~\cite{Bouhon2019,Bouhon2020}. Since a region containing no nodes has $\chi=0$, a non-zero value of $\chi$ indicates a topological obstruction to the possibility of the nodes gapping out in this region. Around each node, a non-zero Euler class $\chi$ manifests itself as a winding $w=2\chi$ in the two band subspace carrying the node. The braiding of nodes in reciprocal space therefore provides a means through which nodes carrying higher winding numbers could be realised in real materials. Importantly, such multi-gap phases are increasingly being related to novel physical effects. Examples include monopole-anti-monopole generation~\cite{Unal2020} that have been seen in trapped-ion experiments~\cite{zhao2022observation} or novel anomalous phases~\cite{AnEuler}, and are increasingly gaining attention in different contexts that range from phononic systems and cold atom simulators to acoustic and photonic metamaterials~\cite{Jiang1Dexp,park2022nodal,Peng2021,  Guo1Dexp, Kemp_2022,konye2021, Jiang_meron, Sjoqviist_2004, peng2022multi, Jiang2021, subdim, Bouhon_magnetic, zhao2022observation,Park2021, Lange2022, chen2021manipulation}.

One way in which the unusual properties of topological materials can become manifest is in their Andreev reflection characteristics. Andreev reflection is the process through which an electronic excitation in a normal metal can scatter into a hole and/or Cooper pair when incident on the boundary between the normal metal and a superconductor. In the historically well-understood case of a quadratic band dispersion, an incident electron which scatters into a hole is retro-reflected, and travels away from the boundary along the same direction from which it arrived \cite{Blonder1982}. In contrast, in graphene an electron incident on a normal-superconductor boundary can in addition undergo specular Andreev reflection. Such unusual scattering characteristics of the single-particle excitations in graphene can ultimately be attributed to the properties of the nodes (Dirac cones) within the band structure of monolayer graphene. The nearest-neighbour tight-binding model of this material famously exhibits degeneracies at two points $\vb{K}$ and $\vb{K}'$ in the Brillouin zone; low-energy excitations about these points behave like relativistic particles with an approximately linear dispersion relation and, due to the non-zero winding numbers carried by the nodes, these excitations are chiral \cite{Beenakker2006, Beenakker2008}. These properties enable a condensed matter realisation of the Klein paradox, which in some cases can lead to perfect Andreev reflection \cite{Lee2019, Akhmerov2008}. Studies have also shown that Rashba spin-orbit interactions can play an important role in Andreev reflection in graphene \cite{Zhai2014, Razieh2016}.  

In some cases it has been found that the Andreev scattering matrix can be directly related to the topological invariants in the bulk of a material. For example, in topological superconductors with chiral symmetry, the quantum number $Q\in\mathbb{Z}$ of the BDI class is equal to the trace of the Andreev S-matrix $r_{\text{he}}$ \cite{Tewari2011, Diez2012, Fulga2011}. Moreover, in certain topological superconductors the interaction of Majorana bound states with electronic excitations can have a significant impact on the Andreev reflection characteristics \cite{Tanaka2009, James2014, Choy2014, San-Jose_2013, Linder2010}. In such materials the differential conductance curves depend on the number of vortices present, and differ in the cases that this number is even and odd \cite{Law2009, Sun2016, Fu2012}. The investigation of Andreev reflection in Weyl semimentals has also shown that the (pseudo-)spin structure of the Hamiltonian in the vicinity of a band node can strongly influence the directional dependence of Andreev reflection \cite{Uchida2014, Chen_2013, Feng2020}.

Motivated by the unusual Andreev reflection properties of topological materials as described above, in this work we explore low-energy Andreev reflection in Euler materials. To do so, we make use of a simple model of a node which carries a generic integer winding number $n\geq 0$ since, as previously mentioned, this is precisely the property exhibited by Euler materials around a band node. (Note that graphene has band nodes with winding numbers $w=\pm 1$; our results agree with previous findings in this limit.) We find that, when $n$ is even, the ability of an electron to scatter into a hole is strongly suppressed at all angles, while for $n$ odd the probability of Andreev reflection at normal incidence is always exactly unity. Moreover, for $n$ even the degree of suppression increases with the strength of an externally applied gate potential $U_0$ (though this effect is less pronounced for large $n$). The differential conductance curves resulting from this behaviour are distinctive and could be measured experimentally and used as an indicator of the presence of Euler topology.

\section{Model}
We now describe and make use of a model to investigate the properties of Andreev reflection in topological Euler materials. Although the braiding of band nodes carrying non-Abelian frame charges requires the Bloch Hamiltonian to have three or more bands, we can obtain an effective description of the low energy physics around any given node by projecting the Hamiltonian to the two-band subspace containing the degeneracy. In this space, a non-zero Euler class is manifested as a winding of the Bloch Hamiltonian around the node. More precisely, the Euler class of a generic real two-band Bloch Hamiltonian in two dimensions
\begin{equation}\label{eq:ham}
H(\mathbf{k}) = a(\vb{k})\mathbbm{1}+r(\mathbf{k}) \cos(\theta(\mathbf{k}))\sigma_x + r(\mathbf{k}) \sin(\theta(\mathbf{k}))\sigma_z,
\end{equation}
where $\mathbf{k} = (k_x, k_y)\in \text{B.Z.}$, is given by $\chi = -w/2$, where the total winding number
\begin{equation}
w = \frac{1}{2 \pi}\int_{\cup_i \partial \mathcal{D}_i} \dd\mathbf{k}\cdot\grad\theta(\mathbf{k}) \in \mathbb{Z}
\end{equation}
and the region $\mathcal{D}_i\subset \text{B.Z.}$ contains the $i^{\text{th}}$ node of $H$. More generally, $\chi = \int_\mathcal{D} Eu-\oint_{\partial\mathcal{D}} a$, where the Euler form $Eu$ and the connection $a$ may be computed from the Berry-Wilczek-Zee connection $\mathcal{A}^{ij} = \ii\bra{u^i}\dd\ket{u^j}$~\cite{wu2019non,Bouhon2020, peng2022multi, Bouhon2020,Ahn2019}.

With regard to the above Hamiltonian we note previous insighful reports that considered the above Hamiltonian in the context of stable winding and band degeneracies~\cite{Montambaux_2018} as well as its use as the simplest ansatz supporting a finite Euler class~\cite{Ahn2019}. The point from a multi-gap perspective~\cite{Bouhon2019,Bouhon2020, peng2022multi,Unal2020} is however that the nodes formed by a two-band subspace on an isolated patch of the Brillouin zone are stable as long as these bands remain disconnected from the other bands of the many-band context. Indeed, to induce a finite Euler class in a lattice model one necessarily needs a multi-gap structure, as can be directly analyzed \cite{Bouhon2019} also in agreement with general homotopical arguments~\cite{Bouhon2020,ChiralHeirs}, and perform a braid between the nodes of adjacent energy gaps. We stress that this general multi-gap perspective sheds light on addressing physical features as recently discovered~\cite{Unal2020,peng2022multi, AnEuler, chen2021manipulation, Peng2021}.

A minimal model of a single node with winding number $w=n$, which for simplicity we choose to be at the origin $\vb{k}=\vb{0}$, is given by setting $\theta(\mathbf{k}) = n \arg(z)=\arg(z^n)$ and $r(\mathbf{k}) = |z|^n= |\mathbf{k}|^n$, where $z = k_x+\ii k_y$. This makes the components of $H$ homogeneous polynomials in $k_x, k_y$, since it then follows that
\begin{subequations}\label{eq:PQ}
\begin{align}
r\cos(\theta) =&{} \Re\{(k_x+\ii k_y)^n\}=: P_n^+(k_x, k_y),\\
r\sin(\theta) =&{} \Im\{(k_x+\ii k_y)^n\}=: P_n^-(k_x, k_y).
\end{align}
\end{subequations}
The first few polynomials are $P_1^+ = k_x, P_2^+ = k_x^2 - k_y^2, P_3^+ = k_x^3 -3k_x k_y^2$ and  $P_1^- = k_y, P^-_2 = 2 k_x k_y, P^-_3 = 3k_x^2 k_y -k_y^3$. In particular, when $n=1$ the Hamiltonian is $H_1 = k_x \sigma_x+k_y \sigma_z$, which is related to the low-energy graphene Hamiltonians $H_{\pm\vb{K}} = k_x \sigma_x\pm k_y \sigma_y$ via a unitary transformation. 

Although the Euler class of the entire Brillouin zone must be quantised to an integer, it is possible to have isolated nodes with arbitrary winding numbers. For this reason, in the following we will allow $n$ to be any integer rather than restricting it to be even.

\begin{figure}
  \centering
  \includegraphics[width=0.235\textwidth]{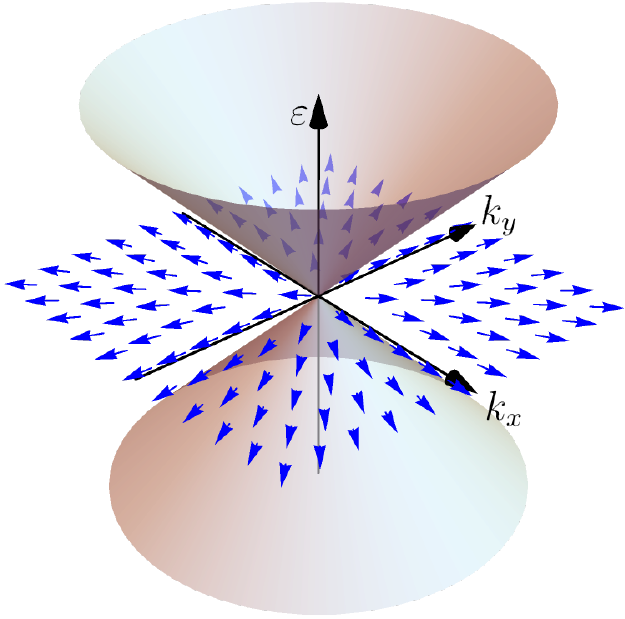}
  \includegraphics[width=0.235\textwidth]{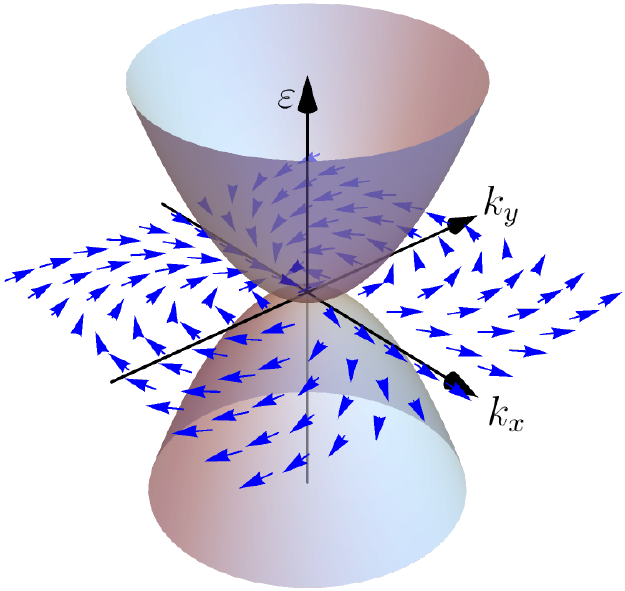}
  \caption{\small  Low energy excitation spectrum of the Hamiltonian $H_n(\vb{k})$ of Eq. \ref{eq:ham} for $n=1$ (left) and $n=2$ (right). The field in the $k_x$-$k_y$ plane displays the vector $(\cos\theta(\vb{k}), \sin\theta(\vb{k}))$, which winds $n$ times around the origin.}
\end{figure}

\begin{figure}
  \centering
  \includegraphics[width=0.48\textwidth]{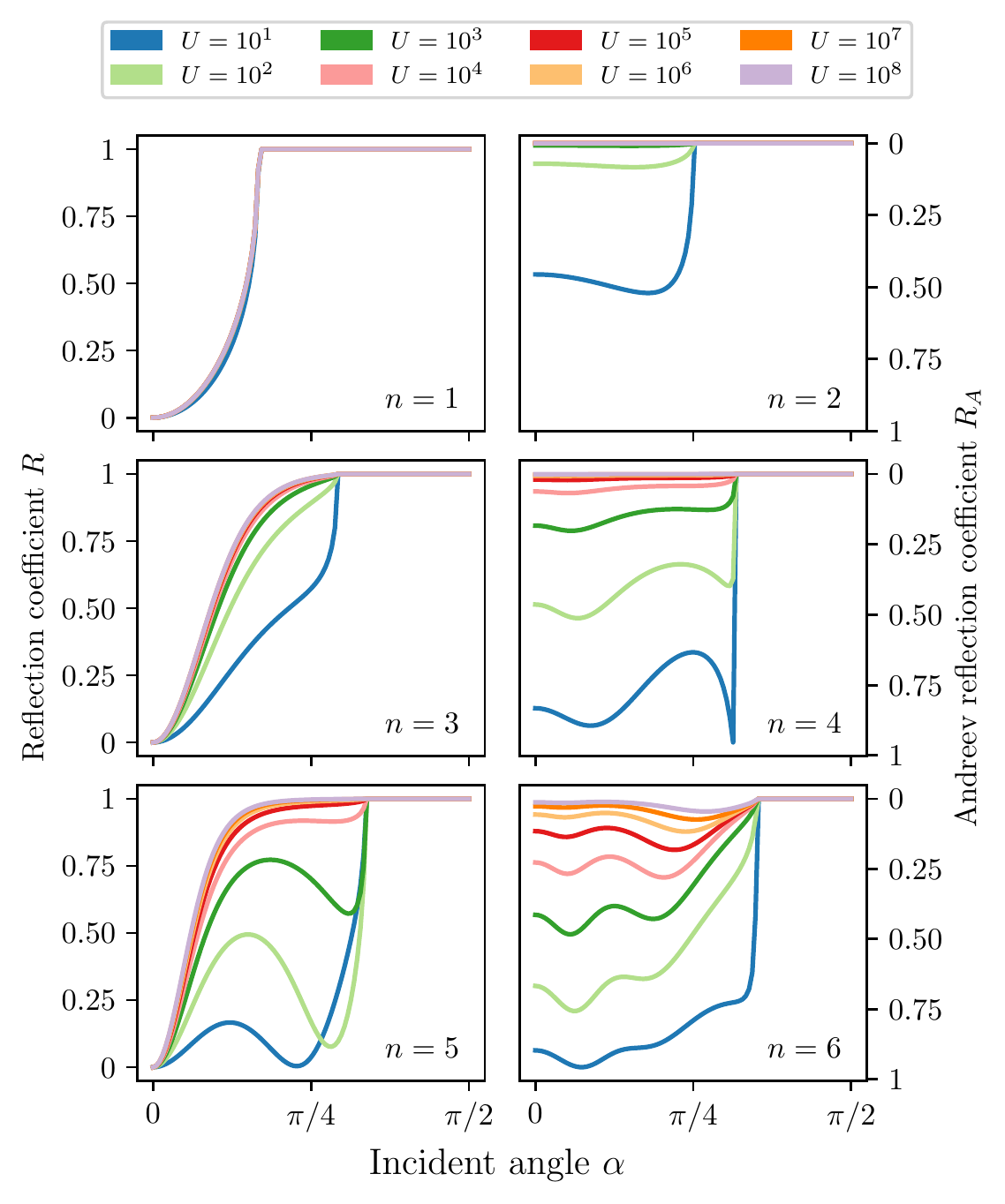}
  \caption{\small  Variation of the reflection and Andreev reflection coefficients $R=|r|^2$ and $R_{\text{A}}=|r_{\text{A}}|^2$ with incident electron angle and applied potential $U$, with $\varepsilon=0.5$, $\Delta_0=1$, and $\EF=1.5$. The energy $\varepsilon$ of the incoming electron is here less than the superconducting gap, $\varepsilon<\Delta_0$, so the sum of the reflection and Andreev reflection coefficients $R$ and $R_{\text{A}}$ is exactly equal to 1.}
  \label{fig:R}
\end{figure}

We now suppose that we have a material which contains a node with a generic winding number $n$ within its bulk band structure; from now on, we will refer to this as an Euler material. If, in addition, a position dependent superconducting pairing potential $\Delta(\vb{x})$ is induced in the material, for example via the proximity effect, then the quasiparticle excitations in the system can be described by a Bogoliubov-de Gennes (BdG) Hamiltonian of the form
\begin{equation}\label{eq:BdG1}
H_{\text{BdG}, n}(\vb{k}) = \mqty( H_n(\vb{k})+U(\vb{x}) -E_{\text{F}} & \Delta(\vb{x}) \\ \Delta(\vb{x})^{\dagger} & E_{\text{F}} - H_n(\vb{k})-U(\vb{x})),
\end{equation}
where $H_n(\vb{k})=P^+_n(\vb{k})\sigma_x+P^-_n(\vb{k})\sigma_z$ is the two-band Euler Hamiltonian with winding number $n$ described above. In Eq.\,(\ref{eq:BdG1}) we have also allowed for the possibility of an externally applied electrostatic potential $U(\vb{x})$. When the pairing and electrostatic potentials vary as a function of position, the momentum $\vb{k}$ should be interpreted as a derivative in real space, and the excitations of the system may be determined by solving the PDE
\begin{equation}\label{eq:BdG}
H_{\text{BdG}, n}(-\ii\partial_{\vb{x}})\psi(\vb{x}) = \varepsilon\psi(\vb{x})
\end{equation}
for positive eigenvalues $\varepsilon\geq 0$, subject to appropriate boundary conditions.

Suppose now that a normal Euler material fills the semi-infinite plane $x<0$, while in the region $x>0$ both the pairing and electrostatic potentials are non-zero and the material is superconducting. In particular, suppose that $\Delta$ and $U$ are uniform in each of these regions,
\begin{align}
\Delta(x, y) &= \begin{cases} \Delta_0 \ee^{\ii\phi}& x>0 \\ 0 & x<0\end{cases}\\
U(x, y) &= \begin{cases} -U_0 & x>0 \\ 0 & x<0\end{cases}.
\end{align}
When an electron propagating in the normal region is incident on the boundary $x=0$, it may scatter into an electron or a hole in the normal region, or a Cooper pair in the superconducting region, each with a certain probability. To determine the amplitudes for these various processes, it is necessary to solve the real-space BdG equation in the normal and superconducting regions. By finding the eigenstates of this equation in the normal and superconducting regions, and then matching these solutions at the boundary $x=0$, we determine the reflection and transmission coefficients. The results of this calculation are shown in the Appendix.

\begin{figure}
  \centering
  \includegraphics[width=0.45\textwidth]{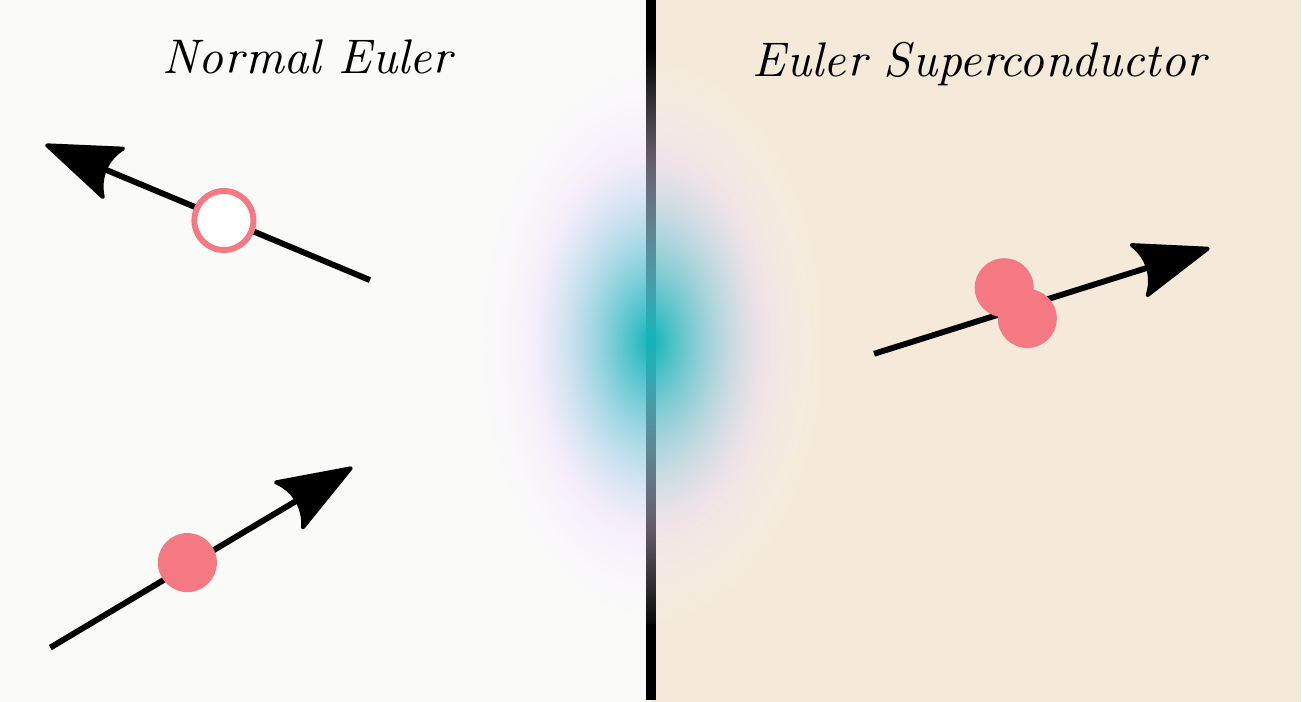}
  \caption{\small  In addition to the propagating wave solutions that describe electron- and hole- like quasiparticle excitations inside the bulk of the normal and superconducting Euler materials, when the winding number is $n$ there exist $2(n-1)$ modes localised at the boundary.}
\end{figure}

In Fig. \ref{fig:R} the probability $R=|r|^2$ for an electron to reflect off the boundary is shown as a function of the incident angle $\alpha$, the winding number $n$, and the strength $U_0$ of the applied electrostatic potential. The energy $\varepsilon$ of the incoming electron is here less than the superconducting gap, $\varepsilon<\Delta_0$, so the sum of the reflection and Andreev reflection coefficients $R$ and $R_{\text{A}}$ is exactly equal to 1. This is because the excitation energy $\varepsilon$ is less than the energy $\Delta_0$ required to create a Cooper pair out of the vacuum, so the electron can scatter only into states on the normal side. Above the critical angle
\begin{equation}
\alpha_{\text{c}} = \arcsin(\left[\frac{|\varepsilon-\EF|}{\varepsilon+\EF}\right]^{1/n})
\end{equation}
there are no hole states available for the electron to scatter into, and the probability of Andreev drops to zero (equivalently, $R=1$). We note that, for fixed $\varepsilon$, $\EF$ the angle $\alpha_{\text{c}}$ increases monotonically with $n$, so that the range of angles over which Andreev reflection can take place is larger for greater values of $n$.

\begin{figure}
  \centering
  \includegraphics[width=0.48\textwidth]{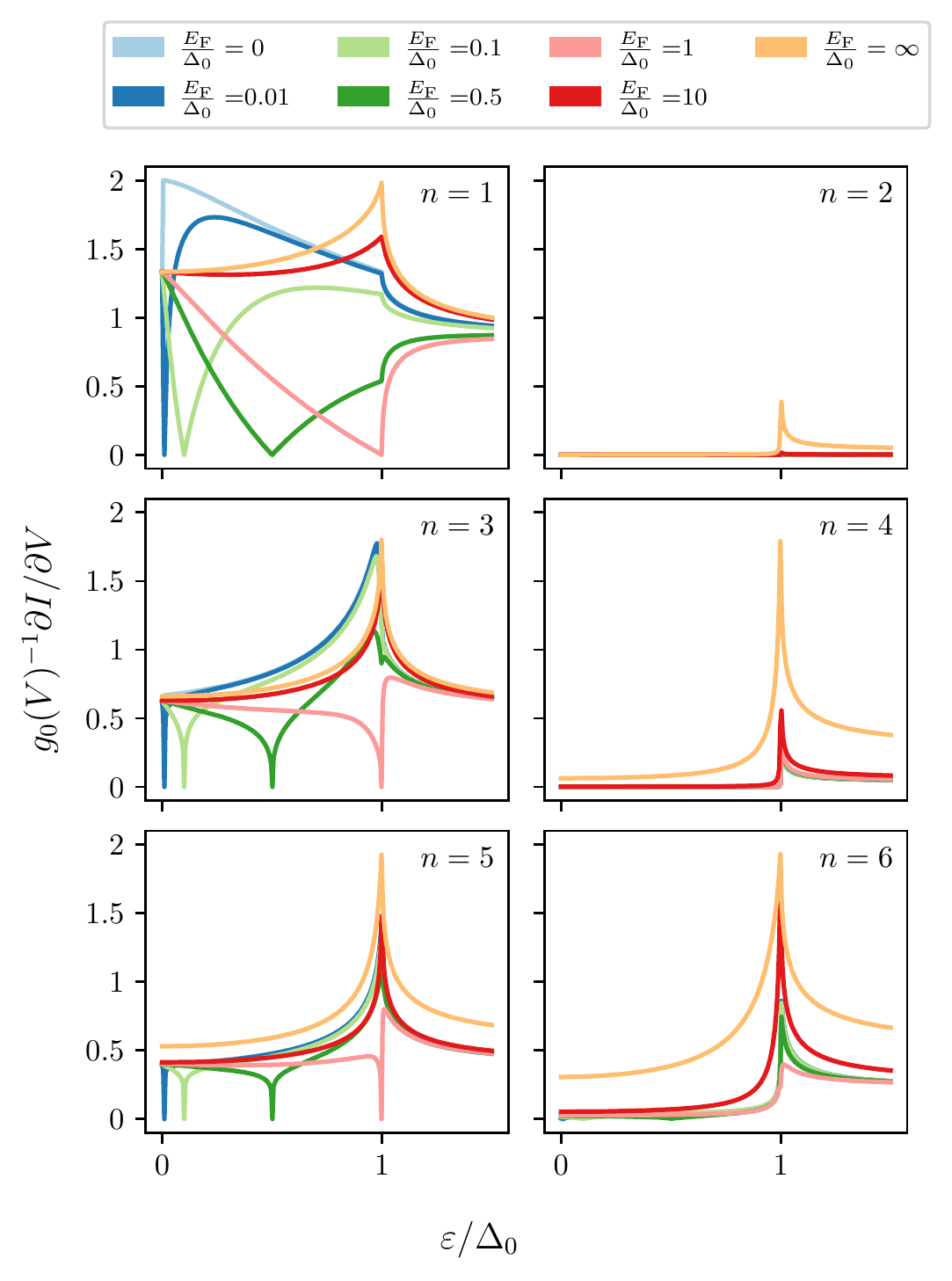}
  \caption{\small  Variation of the differential conductance $\partial I/\partial V$ with the Fermi and excitation energies $E_{\text{F}}$ and $\varepsilon$ with a large external gate potential $U_0 = 10^8$.}
  \label{fig:DiffCond}
\end{figure}

Interestingly, the form of the angular dependence of the reflection probability is qualitatively different when the winding number $n$ is even or odd. For $n$ odd, the probability for Andreev reflection is always exactly 1 at normal incidence ($\alpha=0$), independent of the strength of the potential $U_0$. As $U_0$ is varied, the value of $R$ increases uniformly across $\alpha$. However, even for large changes in $U_0$, the change effected in $R$ is small. On the other hand, for $n$ even the value of $R$ is not fixed to zero at $\alpha =0$, i.e. there is a non-zero probability for the electron to reflect as an electron even at normal incidence. Moreover, $R$ shows a significant variation with the applied potential: as $U_0$ is increased, the value of $R$ increases and tends towards the value $R(\alpha)=1$ as $U_0\to\infty$. The speed at which $R$ increases depends on the magnitude of $n$: for larger $n$, the suppression of $R_{\text{A}}$ is smaller and it requires a stronger potential to send $R\to 1$. These unusual features are qualitatively reproduced in the super-gap region $\varepsilon>\Delta_0$, though in this latter case it no longer holds that $R+R_{\text{A}}=1$.

Another interesting feature of the problem is that, in addition to the propagating wave states that, far into the bulk, represent the usual electron- and hole- excitations,  for $n>1$ there are $2(n-1)$ evanescent solutions that are localised at the boundary $x=0$. Though they do not influence the properties of the propagating wave states away from the boundary, these modes nonetheless make a significant contribution to the dynamics of scattering. Moreover, since these boundary modes exist only for $n>1$, they are a signature of the higher bulk winding number of the Euler material.

The differential conductance of the interface may be calculated using the Blonder-Tinkham-Klapwijk formalism as described in \cite{Blonder1982}: 
\begin{equation}
\frac{1}{g_0(V)} \pdv{I}{V} = \int_0^{\pi/2} \dd\alpha \left[1-R(\alpha)+R_{\text{A}}(\alpha)\right]\cos\alpha.
\end{equation}
The results of this computation for $n=1,\ldots, 6$ are shown for a range of Fermi and excitation energies in Fig. \ref{fig:DiffCond}. Note that, as it must, the case $n=1$ agrees with Fig. 4 of \cite{Beenakker2006}. Most notable however is the result for $n=2$, which shows highly suppressed differential conductance over a wide range of excitation energies; indeed, it is only notably different from zero when the Fermi energy is very large compared to the superconducting gap and the excitation energy is slightly above this same energy $\Delta_0$. 

It is qualitatively clear from Fig. \ref{fig:DiffCond} that the parity dependence displayed in the Andreev reflection in an Euler material (Fig. \ref{fig:R}) is also manifest in the physically measurable quantities $\partial I/\partial V$. For example, on the $n=3$ graph in  Fig. \ref{fig:DiffCond} it can be seen that the magnitude of the differential conductance tends towards  approximately the same value in the limits $\varepsilon\to 0$ and $\varepsilon\gg\Delta_0$. In contrast, the differential conductance in the case $n=6$ tends towards to quite different values on either side of the line $\varepsilon=\Delta_0$.

The plots shown in Fig. \ref{fig:DiffCond} are for a very large applied gate potential $U_0=10^8$. For $U_0\lesssim 10^4$ the corresponding plots are qualitatively similar, but the suppression in the case of $n$ even is less severe.

\section{Conclusion}
We have demonstrated that the parity of the winding number of a band node has a strong influence on the scattering properties of the quasiparticle excitations at the boundary between a normal and a superconducting region. The suppression of the Andreev reflection coefficients in Euler materials with nodes carrying even winding numbers leads to a clear signature in the differential conductance curves that could in principle be probed experimentally.

\section{Acknowledgements}
A.~S.~M. is funded by an EPSRC PhD studentship (Project reference 2606546).  A.~B. was funded by a Marie-Curie fellowship, grant no. 101025315. R.~J.~S acknowledges funding from a New Investigator Award, EPSRC grant EP/W00187X/1, as well as Trinity college, Cambridge. The authors would like to thank X. Feng for helpful discussions on Andreev reflection, and for reading and providing feedback on a draft of the manuscript.

\bibliography{export.bib}

\appendix
\section{The polynomials $P_n^{\pm}(z_1, z_2)$}
In Eq. \ref{eq:PQ} we defined two sets of polynomials $P_n^{\pm}(k_x, k_y)$ as the real and imaginary parts of the complex expression $(k_x+\ii k_y)^n$. While this definition works perfectly well when both arguments $k_x$ and $k_y$ are real, as is the case for propagating wave solutions, when considering localised bound states the wavevectors are generically complex; in this case the definition given above is not sufficient. For complex arguments $z_1, z_2\in\mathbb{C}$ we extend the above definition of the polynomials $P^\pm_n$ to 
\begin{subequations}
\begin{align}
P_n^{+}(z_1, z_2) ={}& \frac{1}{2}\left[(z_1+\ii z_2)^n + (z_1-\ii z_2)^n \right],\\
P_n^{-}(z_1, z_2) ={}& \frac{1}{2\ii}\left[(z_1+\ii z_2)^n - (z_1-\ii z_2)^n \right]
\end{align}
\end{subequations}
or, equivalently,
\begin{subequations}
\begin{align}
P_n^{+}(z_1, z_2) ={}& \sum_{k=0,\, \text{even}}^n\binom{n}{k}(-1)^{k/2}z_1^{n-k}z_2^k,\\
P_n^{-}(z_1, z_2) ={}& \sum_{k=1,\, \text{odd}}^n\binom{n}{k}(-1)^{(k-1)/2}z_1^{n-k}z_2^k.
\end{align}
\end{subequations}
Note that $P_n^{+(-)}(z_1, z_2)$ is equal to the real (imaginary) part of the complex number $(z_1+\ii z_2)^n$ only when both $z_1$ and $z_2$ are real. For instance if $a, b, c\in\mathbb{R}$ then $P_2^+(a+\ii b, c)=(a+\ii b)^2-c^2\neq \Im[(a+\ii b+\ii c)^2]= a^2-(b+c)^2$.

\section{Derivation of scattering coefficients}

In the normal and superconducting regions $x<0$ and $x>0$ we search for wave-like scattering solutions $\psi = \exp(\ii k_x x +\ii k_y y) (u, v)$ of Eq. \ref{eq:BdG} . In each case this leads to a particular eigenvalue equation for the excitation energies $\varepsilon$, and for the vectors $u, v$.

\subsection{Normal side}
On the normal side the pairing potential $\Delta$ vanishes and the gate potential $U=0$. The BdG equation to be solved is therefore
\begin{equation}
\mqty( H_n(\vb{k}) -E_{\text{F}} & 0\\ 0 & E_{\text{F}} - H_n(\vb{k}))\mqty(u \\ v) = \varepsilon \mqty(u\\ v),
\end{equation}
where the excitation energy $\varepsilon>0$. The characteristic equation of this block diagonal matrix factorises into the product of the characteristic equations of each block, so the eigenstates are divided into electronic states (with $u\neq 0$, $v=0$), and hole states (with $u=0$, $v\neq 0$). Explicit expressions for all of these states are as follows:
\begin{subequations}
\begin{align}
\psi^{\text{e}+}_n &= \frac{\ee^{\ii k_{x; n, 0}^{\text{(N, e)}} x}}{\sqrt{v_n^{\text{e}}}}\mqty(\sin(\frac{n\alpha}{2}) + \cos(\frac{n\alpha}{2})\\  - \sin(\frac{n\alpha}{2})+\cos(\frac{n\alpha}{2}) \\0 \\0) \label{eq:Ne0p}\\
\psi^{\text{e}-}_n &= \frac{\ee^{-\ii k_{x; n, 0}^{\text{(N, e)}} x}}{\sqrt{v_n^{\text{e}}}}\mqty(\sin(\frac{n\alpha}{2}) +(-1)^{n+1} \cos(\frac{n\alpha}{2})\\  (-1)^{n+1}\sin(\frac{n\alpha}{2}) -\cos(\frac{n\alpha}{2}) \\0 \\0)\label{eq:Ne0m}\\
\psi^{\text{e}}_{n, q} &= \ee^{\ii k_{x; n, q}^{\text{(N, e)}} x}\mqty(s_{n, q}^{\text{(N, e)}}(\alpha)+1 \\ c_{n, q}^{\text{(N, e)}}(\alpha) \\ 0 \\ 0) \label{eq:Nenq}\\
\psi^{\text{h}+}_n &= \frac{\ee^{\ii k_{x; n, 0}^{\text{(N, h)}} x }}{\sqrt{ v_n^{\text{h}}}}\mqty(0\\ 0 \\ \sin(\frac{n\alpha'}{2}) - S^{n+1}\cos(\frac{n\alpha'}{2}) \\ S^{n+1}\sin(\frac{n\alpha'}{2})+ \cos(\frac{n\alpha}{2}) )\label{eq:Nh0p}\\
\psi^{\text{h}-}_n &= \frac{\ee^{-\ii k_{x; n, 0}^{\text{(N, h)}} x}}{\sqrt{v_n^{\text{h}}}}\mqty(0\\ 0 \\ \sin(\frac{n\alpha'}{2}) - (-S)^{n+1}\cos(\frac{n\alpha'}{2}) \label{eq:Nh0m}\\ -(-S)^{n+1}\sin(\frac{n\alpha'}{2})-\cos(\frac{n\alpha}{2}) ) \\
\psi_{n, q}^{\text{h}} &=  \ee^{\ii k_{x; n, q}^{\text{(N, h)}} x}\mqty(0 \\ 0 \\ s_{n, q}^{\text{(N, h)}}(\alpha') -S \\ c_{n, q}^{\text{(N, h)}}(\alpha')).\label{eq:Nhnq}
\end{align}
\end{subequations}
where in Eqs. \ref{eq:Nenq} and \ref{eq:Nhnq} the index $q=1, 2, \ldots, n-1$, and factors of $\ee^{\ii k_y y}$ common to all terms have been omitted since they cancel in the final equations. In the above expressions the letters `e' and `h' indicate whether the state corresponds to an electron or a hole, and the label `N' indicates that the wavevectors are valid in the normal region. The states $\psi^{\text{e}, +}_n$ and $\psi^{\text{h}, +}_n$ describe excitations propagating in the positive $x$-direction, while the states $\psi^{\text{e}, -}_n$ and $\psi^{\text{h}, -}_n$ propagate in the negative $x$-direction. The electron and hole wavevectors for these states are real and given by
\begin{subequations}
\begin{align}
k_{x; n, 0}^{\text{(N, e)}} &= (\varepsilon+\EF)^{1/n} \cos\alpha,\\
k_{x; n, 0}^{\text{(N, h)}} &= S|\varepsilon-\EF|^{1/n}\cos\alpha',
\end{align}
where $S = \sgn(\varepsilon-E_{\text{F}})$. On the other hand, the wavevectors 
\begin{align}
k_{x; n, q}^{\text{(N, e)}} &= \bar{S}_{n, q} (\varepsilon+\EF)^{1/n}\sqrt{r_{n, q}(\alpha)}\,\ee^{\ii \phi_{n, q}(\alpha)/2}\\
k_{x; n, q}^{\text{(N, h)}} &= \bar{S}_{n, q}|\varepsilon-\EF|^{1/n}\sqrt{r_{n, q}(\alpha')}\ee^{\ii\phi_{n, q}(\alpha')/2}
\end{align}
\end{subequations}
for the states $\psi^{\text{e}}_{n, q}$ and $\psi^{\text{h}}_{n, q}$ are complex, so these states are evanescent and describe modes localised at the boundary. The signs 
\begin{equation}
\bar{S}_{n, q} = \begin{cases} -1 & q<\frac{n}{2}\\
1 & \text{otherwise}. \end{cases}
\end{equation}
ensure that these wavevectors all have positive imaginary parts so  that the states decay to zero in the region $x<0$, rather than blowing up as $x\to -\infty$. The existence of these $n-1$ additional states can be traced back to the fact that the BdG Hamiltonian depends on $|\vb{k}|^n$, which in real space translates to an $n^{\text{th}}$ order differential equation - precisely $n$ states for each energy.

The speeds of propagation of the electron and hole states in the $x$-direction are respectively given by
\begin{subequations}
\begin{align}
v_n^{\text{e}}&= n (\varepsilon+\EF)^{\frac{n-1}{n}} \cos\alpha,\\
v_n^{\text{h}}&=n|\varepsilon-E_{\text{F}}|^{\frac{n-1}{n}} \cos\alpha';
\end{align}
\end{subequations}
the states in Eqs. \ref{eq:Ne0p}, \ref{eq:Nh0m}, \ref{eq:Nh0p}, and \ref{eq:Ne0m} have been normalised by their respective speeds to ensure that they carry a single particle. (Since the states \ref{eq:Nenq} and \ref{eq:Nhnq} are not propagating their normalisation is unimportant so has been omitted). 

The angle $\alpha$ at which the incident electron propagates with respect to the $x$-axis may be expressed in terms of the conserved quantities $k_y$ and $\varepsilon$ as 
\begin{subequations}
\begin{equation}
\alpha = \arcsin(\frac{k_y}{\left(\varepsilon+E_{\text{F}}\right)^{1/n}} ),
\end{equation}
and the corresponding angle for the propagating hole state is
\begin{equation}
\begin{split}
\alpha'&= S\arcsin(\frac{k_y}{|\varepsilon-E_{\text{F}}|^{1/n}}) \\
&= S\arcsin(\left(\frac{\epsilon+\EF}{|\varepsilon-E_{\text{F}}|^{1/n}}\right)\sin\alpha). \\
\end{split}
\end{equation}
\end{subequations}
The remaining terms are
\begingroup
\allowdisplaybreaks
\begin{subequations}
\begin{align}
c_{n, q}^{\text{(N, e)}} &= P_n^+\left[\bar{S}_{n, q}\sqrt{r_{n, q}(\alpha)}\,\ee^{\ii\phi_{n, q}(\alpha)/2}, \sin(\alpha) \right]\\
s_{n, q}^{\text{(N, e)}} &= P_n^-\left[\bar{S}_{n, q}\sqrt{r_{n, q}(\alpha)}\,\ee^{\ii\phi_{n, q}(\alpha)/2}, \sin(\alpha) \right]\\
c_{n, q}^{\text{(N, h)}} &= P_n^+\left[\bar{S}_{n, q}\sqrt{r_{n, q}(\alpha')}\ee^{\ii\phi_{n, q}(\alpha')/2}, S\sin(\alpha')\right]\\
s_{n, q}^{\text{(N, h)}} &= P_n^-\left[\bar{S}_{n, q}\sqrt{r_{n, q}(\alpha')}\ee^{\ii\phi_{n, q}(\alpha')/2}, S\sin(\alpha')\right]
\end{align}
\end{subequations}
where the polynomials $P^\pm(z_1, z_2)$ are defined in Eq. \ref{eq:PQ}, and
\begin{subequations}\label{eq:rphi}
\begin{align}
r_{n, q}(\beta) ={}& \sqrt{\left(\cos(\frac{2\pi q}{n})-\sin^2(\beta)\right)^2 +\sin^2\left(\frac{2\pi q}{n}\right)},\\
\phi_{n, q}(\beta) ={}& \arctan\left\{\cos(\frac{2\pi q}{n})-\sin^2\beta, \sin(\frac{2\pi q}{n}) \right\}.
\end{align}
\end{subequations}
The function $\arctan(x, y)$ is here defined to give the angle subtended between the point $(x, y)$ and the positive $x$-axis, and to have a single branch cut along the line $y=0$, $x<0$.

\subsection{Superconducting side}
On the superconducting side the pairing potential $\Delta\neq 0$ and the gate potential $U=-U_0$ is finite, so the BdG equation becomes
\begin{equation}
\mqty( H_n(\vb{k})+U_0 -E_{\text{F}} & \Delta_0 \ee^{\ii \phi} \\ \Delta_0 \ee^{- \ii \phi} & E_{\text{F}} - H_n(\vb{k})-U_0)\mqty(u \\ v) = \varepsilon \mqty(u\\ v).
\end{equation}
The characteristic equation now no longer factorises, indicating that the basic quasiparticle excitations of the system consist of coherent superpositions of electrons and holes together. We now list all of the scattering states on the superconducting side:
\begin{subequations}
\begin{align}
\varphi^{\text{e}+}_n ={}& \ee^{\ii k_{x; n, 0}^{\text{(S, e)}} x}\mqty(\ee^{b}\left(\sin(\frac{n\gamma}{2})+\cos(\frac{n\gamma}{2})\right) \\ \ee^{b}\left(-\sin(\frac{n\gamma}{2})+\cos(\frac{n\gamma}{2})\right) \\ \ee^{-\ii\phi} \left(\sin(\frac{n\gamma}{2})+\cos(\frac{n\gamma}{2})\right) \\ \ee^{-\ii\phi}\left(-\sin(\frac{n\gamma}{2})+\cos(\frac{n\gamma}{2})\right) ) \\
\varphi^{\text{e}-}_n ={}& \ee^{-\ii k_{x; n, 0}^{\text{(S, e)}}}  \mqty(\ee^{b}\left(\sin(\frac{n\gamma}{2})+(-1)^{n+1}\cos(\frac{n\gamma}{2})\right) \\ \ee^{b}\left((-1)^{n+1}\sin(\frac{n\gamma}{2})-\cos(\frac{n\gamma}{2})\right) \\ \ee^{-\ii\phi}\left(\sin(\frac{n\gamma}{2})+(-1)^{n+1}\cos(\frac{n\gamma}{2})\right) \\  \ee^{-\ii\phi}\left((-1)^{n+1}\sin(\frac{n\gamma}{2})-\cos(\frac{n\gamma}{2})\right) )\\
\varphi_{n, q}^{\text{e}} ={}& \ee^{\ii k_{x; n, q}^{\text{(S, e)}} x} \mqty(\ee^b\left(s_{n, q}^{\text{(S, e)}}+1\right) \\ \ee^b c_{n, q}^{\text{(S, e)}} \\ \ee^{-\ii\phi}(s_{n, q}^{\text{(S, e)}}+1) \\ \ee^{-\ii\phi} c_{n, q}^{\text{(S, e)}})\\
\varphi^{\text{h}+}_n ={}& \ee^{\ii k_{x; n, 0}^{\text{(S, h)}}x}  \mqty(\ee^{-b}\left(\sin(\frac{n\gamma'}{2})+(-1)^{n+1}\cos(\frac{n\gamma'}{2})\right) \\ \ee^{-b}\left((-1)^{n+1}\sin(\frac{n\gamma'}{2})-\cos(\frac{n\gamma'}{2})\right) \\ \ee^{-\ii\phi}\left(\sin(\frac{n\gamma'}{2})+(-1)^{n+1}\cos(\frac{n\gamma'}{2})\right) \\  \ee^{-\ii\phi}\left((-1)^{n+1}\sin(\frac{n\gamma'}{2})-\cos(\frac{n\gamma'}{2})\right) )\\
\varphi^{\text{h}-}_n ={}& \ee^{\ii k_{x; n, 0}^{\text{(S, h)}} x}\mqty(\ee^{-b}\left(\sin(\frac{n\gamma'}{2})+\cos(\frac{n\gamma'}{2})\right) \\ \ee^{-b}\left(-\sin(\frac{n\gamma'}{2})+\cos(\frac{n\gamma'}{2})\right) \\ \ee^{-\ii\phi} \left(\sin(\frac{n\gamma'}{2})+\cos(\frac{n\gamma'}{2})\right) \\ \ee^{-\ii\phi}\left(-\sin(\frac{n\gamma'}{2})+\cos(\frac{n\gamma'}{2})\right) )\\
\varphi_{n, q}^{\text{h}} ={}& \ee^{\ii k_{x; n, q}^{\text{(S, h)}} x} \mqty(\ee^{-b}\left(s_{n, q}^{\text{(S, h)}}+1\right) \\ \ee^{-b} c_{n, q}^{\text{(S, h)}} \\ \ee^{-\ii\phi}(s_{n, q}^{\text{(S, h)}}+1) \\ \ee^{-\ii\phi} c_{n, q}^{\text{(S, h)}})
\end{align}
\end{subequations}
where again $q=1, 2, \ldots, n-1$, and the letter `S' indicates that we are now dealing with the superconducting side. The letters `e' and `h' now indicate whether the excitations are `electron-like' or `hole-like', that is, whether they propagate in the same or the opposite directions to their group velocities. These names are also motivated by the fact that $\varphi^\text{e}\to\psi^\text{e}$ and $\varphi^\text{h}\to\psi^\text{h}$ in the limit that $\Delta_0\to 0$.

Again, there are two real wavevectors 
\begin{subequations}
\begin{align}
k_{x; n, 0}^{\text{(S, e)}} ={}& \left(E_0+\sqrt{\varepsilon^2+\Delta_0^2}\right)^{1/n}\cos\gamma\\
k_{x; n, 0}^{\text{(S, h)}} ={}& - \left(E_0-\sqrt{\varepsilon^2+\Delta_0^2}\right)^{1/n}\cos\gamma'
\end{align}
corresponding to the forwards- and backwards-propagating electron- and hole-like solutions $\varphi^{\text{e}\pm}_n$ and $\varphi^{\text{h}\pm}_n$. The remaining wavevectors  
\begin{align}
k_{x; n, q}^{\text{(S, e)}} ={}& -\bar{S}_{n, q}\left(E_0+\sqrt{\varepsilon^2-\Delta_0^2}\right)^{1/n}\sqrt{r_{n, q}(\gamma)}\,\ee^{\ii\phi_{n, q}(\gamma)/2}\\
k_{x; n, q}^{\text{(S, h)}} ={}& -\bar{S}_{n, q}\left(E_0-\sqrt{\varepsilon^2-\Delta_0^2}\right)^{1/n}\sqrt{r_{n, q}(\gamma')}\,\ee^{\ii\phi_{n, q}(\gamma')/2}
\end{align}
\end{subequations}
are complex, implying that the states $\varphi^{\text{e}}_{n, q}$ and $\varphi^{\text{h}}_{n, q}$ are evanescent waves. The signs $\bar{S}_{n, q}$ now ensure that these states decay to zero as $x\to +\infty$.

The angles at which the propagating states travel are given by
\begin{subequations}
\begin{align}
\gamma ={}& \arcsin\left[\left(\frac{\varepsilon+\EF}{E_0+\sqrt{\varepsilon^2-\Delta_0^2}}\right)^{1/n}\sin\alpha \right],\\
\gamma' ={}& -\arcsin\left[\left(\frac{\varepsilon+\EF}{E_0-\sqrt{\varepsilon^2-\Delta_0^2}}\right)^{1/n}\sin\alpha \right].
\end{align}
\end{subequations}
Finally, we have
\begin{subequations}
\begin{align}
c_{n, q}^{\text{(S, e)}} ={}& P_n^+\left[-\bar{S}_{n, q}\sqrt{r_{n, q}(\gamma)}\,\ee^{\ii\phi_{n, q}(\gamma_n)/2}, \sin\gamma\right]\\
s_{n, q}^{\text{(S, e)}} ={}& P_n^-\left[-\bar{S}_{n, q}\sqrt{r_{n, q}(\gamma)}\,\ee^{\ii\phi_{n, q}(\gamma_n)/2}, \sin\gamma\right]\\
c_{n, q}^{\text{(S, h)}} ={}& P_n^+\left[-\bar{S}_{n, q}\sqrt{r_{n, q}(\gamma')}\,\ee^{\ii\phi_{n, q}(\gamma')/2}, -\sin\gamma'\right]\\
s_{n, q}^{\text{(S, e)}} ={}& P_n^-\left[-\bar{S}_{n, q}\sqrt{r_{n, q}(\gamma')}\,\ee^{\ii\phi_{n, q}(\gamma')/2}, -\sin\gamma'\right],\nonumber
\end{align}
\end{subequations}
where again $P^\pm$ is defined in Eq. \ref{eq:PQ}, and $r_{n, q}, \phi_{n, q}$ are as in Eqs. \ref{eq:rphi}.

\subsection{Matching}
To determine the probability that the incident electron will be scattered into an outgoing electron or hole on the normal side, or an electron- or hole-like quasiparticle in the superconducting side, we propose a trial ansatz on both sides and determine the amplitudes of each state in the solution. The incoming state on the normal side is a sum of a right-moving electron, a left-moving electron, a left-moving hole, and all evanescent states:
\begin{subequations}\label{eq:ansatze}
\begin{align}
\psi_n ={}& \psi_n^{\text{e}+}\ee^{\ii k_{x; n, 0}^{\text{(N, e)}}} +r\psi_n^{\text{e}-}\ee^{-\ii k_{x; n, 0}^{\text{(N, e)}}} +r_{\text{A}}\psi_n^{\text{h}-}\ee^{-\ii k_{x; n, 0}^{\text{(N, h)}}} \nonumber\\
&+\sum_{q=1}^{n-1}\left[a_q \psi_{n, q}^{\text{e}}\ee^{\ii k_{x; n, q}^{\text{(N, e)}}}+b_q \psi_{n, q}^{\text{h}}\ee^{\ii k_{x; n, q}^{\text{(N, h)}}} \right].
\end{align}
On the superconducting side, the ansatz is a sum of an electron-like and a hole-like quasiparticle propagating to the right, and all evanescent states:
\begin{align}
\begin{split}
\varphi_n ={}& t\varphi_n^{\text{e}+}\ee^{\ii k_{x; n, 0}^{\text{(S, e)}}} +t'\varphi_n^{\text{h}+}\ee^{\ii k_{x; n, 0}^{\text{(S, h)}}} \\
&+\sum_{q=1}^{n-1}\left[a_q \varphi_{n, q}^{\text{e}}\ee^{\ii k_{x; n, q}^{\text{(S, e)}}}+b_q \varphi_{n, q}^{\text{h}}\ee^{\ii k_{x; n, q}^{\text{(S, h)}}}\right].
\end{split}
\end{align}
\end{subequations}
(The factor $\ee^{\ii k_y y}$ common to all terms has again been omitted in Eqs. \ref{eq:ansatze}.)
To determine the coefficients, we apply the boundary/matching conditions
\begin{equation}
\partial_x^p\psi_n(x, y)|_{x=0} = \partial_x^p\varphi_n(x, y)|_{x=0}
\end{equation}
for $ p=0, 1, \ldots, n-1$. This set of $n$ conditions on the four-component vectors $\psi, \varphi$ leads to 

\medskip
\begin{widetext}
\centering
\begin{eqnarray}
I_0 \big(\ii k_{x, 0}^{(\text{N, e})}\big)^p\psi_{\text{e}, +} + r \big(-\ii k_{x, 0}^{(\text{N, e})}\big)^p\psi_{\text{e}, -}+r_{\text{A}}\big(-\ii k_{x, 0}^{(\text{N, h})}\big)^p\psi_{\text{h}, -}+\sum_{q=1}^{n-1}\Big[a_q \big(\ii k_{x, q}^{(\text{N, e})}\big)^p \psi_{\text{e}, q}+b_p\big(\ii k_{x, q}^{(\text{N, h})}\big)^p\psi_{\text{h}, q}\Big] \\ \nonumber
= t\big(\ii k_{x, 0}^{(\text{S, e})}\big)^p \phi_{\text{e}, +} + t'\big(\ii k_{x, 0}^{(\text{S, h})}\big)^p \phi_{\text{h}, +}+\sum_{q=1}^{n-1}\Big[c_q\big(\ii k_{x, q}^{(\text{S, e})}\big)^p\phi_{\text{e}, q}+d_q\big(\ii k_{x, q}^{(\text{S, h})}\big)^p\phi_{\text{h}, q}\Big]
\end{eqnarray}
\end{widetext}
for $p=0, 1, \ldots, n-1$; this is a set of $4n$ equations which may be solved to yield the $4n$ coefficients $r, r_\text{A}, a_1, \ldots, d_{n-1}$. Including the normalisation $I_0=1$ allows these equations to be put into the matrix form $AX=v_0$, where 
\begin{align}
X = (&I_0, r, r_\text{A}, a_1, \ldots, a_{n-1}, b_1, \ldots, b_{n-1},\\
&-t, -t', -c_1, \ldots, -c_{n-1}, -d_1, \ldots, -d_{n-1})^T \nonumber
\end{align}
is a $4n+1$-dimensional vector and $A = (v_0, M^T)^T$ is a $(4n+1)\times (4n+1)$ square matrix consisting of a $1\times (4n+1)$ row vector $v_0^T = (1, 0, 0, \ldots, 0)$, and a $4n\times (4n+1)$-dimensional matrix $M$. The matrix $M$ is the column-wise Khatri-Rao product $M=K\star \Psi$ of the $n\times (4n+1)$-dimensional Vandermonde matrix of wavevectors
\begin{equation}
K = \mqty(1 & 1 & \cdots & 1\\
k_{x, 0}^{(\text{N, e})} & -\ii k_{x, 0}^{(\text{N, e})} & \cdots &  k_{x, n-1}^{(\text{S, h})} \\
\left(k_{x, 0}^{(\text{N, e})}\right)^2 & \left(-k_{x, 0}^{(\text{N, e})}\right)^2 &\cdots & \left(\ii k_{x, n-1}^{(\text{S, h})}\right)^2\\
\vdots & & & \vdots \\
\left(k_{x, 0}^{(\text{N, e})}\right)^{n-1} & \left(-k_{x, 0}^{(\text{N, e})}\right)^{n-1} & \cdots & \left(\ii k_{x, n-1}^{(\text{S, h})}\right)^{n-1})
\end{equation}
and the $4\times (4n+1)$-dimensional matrix of states $ \Psi = \mqty(\psi_{\text{e}, +} & \psi_{\text{e}, -} & \psi_{\text{h}, -} & \cdots & \phi_{\text{h}, p}). $
The vector $v_0^T$ is used to set the normalisation $I_0=1$. The matrix $A$ is non-singular so this yields a unique solution for each of the coefficients $r, r_{\text{A}}$, etc. The expressions resulting from the solution of this set of equations are rather involved, so we omit them; they may be obtained using a computer algebra program. Importantly, when the incident angle $\alpha$ is greater than a critical angle 
\begin{equation}
\alpha_{\text{c}} = \arcsin(\left[\frac{|\varepsilon-\EF|}{\varepsilon+\EF}\right]^{1/n}),
\end{equation}
the hole states are no longer valid scattering states. After the final result has been obtained, all of the coefficients $r_{\text{A}}, b_1, \ldots, b_{n-1}$ must therefore be multiplied by $\theta(\alpha - \alpha_{\text{c}})$, where $\theta(x)$ is the Heaviside step function.

\end{document}